\documentclass[epj]{svjour}
\usepackage{graphics}
\usepackage{multirow}
\usepackage{float}

\newcommand\ba{\begin{eqnarray}}
\newcommand\ea{\end{eqnarray}}
\newcommand\nn{\nonumber}
\newcommand{\be}{\begin{equation}}
\newcommand{\ee}{\end{equation}}
\newcommand{\PANDA}{$\rm \overline{P}ANDA$ }
\begin{document}

\title{Feasibility studies of the time-like proton electromagnetic form factor measurements with $\rm \bf\overline{P}ANDA$ at FAIR.}
\author{M.~Sudo\l \inst{1}, M.C.~Mora Esp\'i \inst{2}, E.~Becheva \inst{1,}\thanks{\emph{Present address:} LLR-Ecole polytechnique, 91128 Palai\-seau, France}, J.~Boucher \inst{1}, T.~Hennino \inst{1}, R.~Kunne \inst{1}, D.~Marchand \inst{1}, S.~Ong \inst{1}, B.~Ramstein \inst{1},   
J.~Van~de~Wiele \inst{1}, T.~Zerguerras \inst{1}, F.~Maas \inst{2,3}, B.~Kopf \inst{4},~ M.~Pelizaeus \inst{4}, M.~Steinke \inst{4},  J.~Zhong \inst{4}, \\
and E.~Tomasi-Gustafsson \inst{5,1,}\thanks{\emph{Corresponding author:} etomasi@cea.fr}}

%
%
\institute{Institut de Physique Nucl\'eaire, CNRS/IN2P3 and Universit\'e  Paris-Sud, France  
\and Johannes Gutenberg Universitaet Mainz, 
Institut fuer Kernphysik, 55099 Mainz, Germany 
\and GSI Helmholtzzentrum fuer Schwerionenforschung, GmbH, 64291 Darmstadt, Germany  
\and Ruhr Universitaet Bochum, 44801 Bochum, Germany
\and DSM, IRFU, SPhN, Saclay, 91191 Gif-sur-Yvette, France }
\date{Received: date / Revised version: date}
%
\abstract{
The possibility of measuring the proton electromagnetic form factors in the time-like region at FAIR with the \PANDA detector is discussed. Detailed simulations on signal efficiency for the annihilation of $\bar p +p $ into a lepton pair as well as for the most important background channels have been performed. It is shown that precise measurements of the differential cross section of the reaction $\bar p +p \to e^-+ e^+$  can be obtained in a wide kinematical range. The determination of the ratio ${\cal R}$ of the moduli of 
the electric and magnetic proton form factors will be possible up to a value of momentum transfer squared of $q^2\simeq 14$ (GeV/c)$^2$ with absolute precision from  0.01 to 0.5 (for ${\cal R}\sim 1$). The total 
$\bar p +p\to e^-+e^+$ cross section will be measured up to $q^2\simeq 28$ (GeV/c)$^2$. The results obtained from simulated events are compared to the existing data. Sensitivity to the two photons exchange mechanism is also investigated.
}
%

\authorrunning{M.~Sudo\l, M.C.~Mora Esp\'i...}
\titlerunning{Time-like proton form factors with \PANDA at FAIR}

\maketitle

\section{Introduction}
\label{intro}

The availability of a high intensity antiproton beam up to a momentum of 15 GeV/c at the FAIR facility \cite{FAIR} and of the \PANDA detector offers unique possibilities for new investigations in the field of hadron structure (see \cite{PANDA} for a review). Here we focus on feasibility studies for the determination of the proton electromagnetic form factors (FFs), in the time-like (TL) region \cite{ETGnote}, through the annihilation reaction:
\be 
\bar p + p\to\ell^-+ \ell^+,~  \ell=e,\mu.
\label{eq:eq1}
\ee
The underlying mechanism is assumed to be the exchange of one virtual photon of four momentum squared $q^2$. The sensitivity of the measurement to higher exchanges, which are in principle suppressed, is also investigated. Muons carry the same physical information on the nucleon structure as the electrons, but this work will focus on the electron channel only. Although the measurements of electromagnetic nucleon FFs have been going on since more than fifty years, major progress has been recently achieved in a wide kinematical region, mostly in space-like (SL) through polarized elastic electron proton scattering. The determination of form factors is limited by the steep decrease of the cross section with $q^2$. Moreover in the time-like (TL) region measurements in both channels: $\bar p + p\leftrightarrow \ell^-+ \ell^+$ are scarce and affected by poor statistics. 

The intensity of the antiproton beam, together with the performances of the \PANDA detector, will make possible the determination of FFs up to large $q^2$. 
FFs are extracted from the angular distribution of one of the charged leptons. In reaction (\ref{eq:eq1}), the difficulty of the measurement is related to the hadronic background, mostly annihilation into pions, which is six order of magnitudes larger than the production of a lepton pair. In this paper we report on detailed simulations of the hadronic background and discuss the precision and the significance of the extracted data in a wide kinematical range.

This paper is structured as follows. In sect. \ref{Physics} the interest of measuring proton electromagnetic form factors is briefly recalled, and the present experimental situation both in SL and TL regions is illustrated. In sect. \ref{sec:Simul} simulation studies for both signal and background are reported. In sect.  \ref{sec:Results} the results on the extracted FFs and on their precision are discussed and compared with the existing data. The sensitivity to the two photon exchange mechanism is discussed in sect. \ref{sec:twophoton}. In sect. \ref{sec:Comparison}, the \PANDA expected performances are compared to the ones obtained in previous experiments. The main results are summarized in the conclusions. 
\section{Physics motivation}
\label{Physics}
Hadron electromagnetic FFs describe the internal structure of a particle. Elastic FFs contain information on the hadron ground state, and are traditionally measured using electron hadron elastic scattering, assuming that the interaction occurs through one-photon exchange (OPE). Assuming a Parity and Time invariant theory, a hadron with spin $S$ is described by $2S+1$ independent FFs. Protons and neutrons (spin 1/2 particles) are then characterized by two form factors, an electric $G_E$ and a magnetic $G_M$ which are analytical functions of one kinematical variable $q^2$. 

\subsection{Space-like region}
Elastic electron proton scattering allows to access the SL  region, where FFs are real functions of $Q^2=-q^2>0$. Electromagnetic FFs are determined through the $\epsilon$ dependence of the (reduced) elastic differential cross section, which may be written, in OPE approximation, as \cite{Ro50}:
\ba
\sigma_{red}&(\theta_e,Q^2)&=\left [1+2\displaystyle\frac{E}{M}\sin^2(\theta_e/2)\right ]\displaystyle\frac
{4 E^2\sin^4(\theta_e/2)}{\alpha_e^2\cos^2(\theta_e/2)}\times
\nonumber\\ 
&&
\times\epsilon(1+\tau)\displaystyle\frac{d\sigma}{d\Omega}
=\tau G_M^2(Q^2)+\epsilon G_E^2(Q^2),
\label{eq:sigma}
\ea
$$
\epsilon=[1+2(1+\tau)\tan^2(\theta_e/2)]^{-1},~~\tau=\displaystyle\frac{Q^2}{4M^2},
$$
$\alpha_e=1/137$, is the electromagnetic fine structure constant,  $M$ is the proton mass, $E$ is the incident electron energy and $\theta_e$ is the scattering angle of the outgoing electron.  Measurements of $\sigma_{red}(\theta_e,Q^2)$ at different angles for a fixed value of $Q^2$ allow to extract $G_E(Q^2)$ and $G_M(Q^2)$  from the slope and the intercept
of the linear $\epsilon$ dependence (\ref{eq:sigma}) (Rosenbluth separation).

The existing data on $G_{M}$ are described by a dipole behavior up to the highest 
measured value $Q^2\simeq$ 31 (GeV/c)$^2$ \cite{Ar86}, according to:
\begin{equation}
G_M(Q^2)/\mu=G_d(Q^2),~
G_d(Q^2)=\left (1+Q^2/0.71\right )^{-2},
\label{eq:dipole}
\end{equation}
where $\mu\sim 2.79 $ is the proton anomalous magnetic moment in nuclear magnetons and $Q^2$ is expressed in ({GeV/c})$^2$.
The independent determination of $G_M$ and $G_E$ from the unpolarized $e^-p$ cross section has been obtained up to $Q^2=$ 8.8 (GeV/c)$^2$ \cite{And94},
and gives $G_E\sim G_M/\mu$. Further extraction of $G_M$ \cite{Ar86} is based on this assumption. 

Experimental and theoretical studies have been done since a few decades, but recent interest aroused due to the possibility to reach higher precision and larger values of $q^2$ at high intensity accelerators, using polarized beams, targets and polarimeters in the GeV range \cite{CFP08}. In particular recent measurements of the FF ratio \cite{Jlab}, based on the polarization method \cite{Re68,Re74}, show that the electric and magnetic distributions in the proton are different, contrary to what was earlier assumed. The $Q^2$ dependence of $G_E$ and $G_M$, deduced from polarization experiments $p(\vec e,e)\vec p$ differs from (\ref{eq:dipole}). The FF ratio shows a linear deviation from a constant, which can be parametrized as \cite{GEp3}:
\begin{equation}
\mu G_E/G_M= 1.059 -0.143 ~Q^2~[(\mbox{GeV/c})^2] \mbox{~for~}  Q^2
\ge 0.4,
\label{eq:polar}
\end{equation}
up to at least $Q^2$=5.8 (GeV/c)$^2$. Polarization measurements have recently been extended up to $Q^2$=8.5 (GeV/c)$^2$ by the GEP collaboration, at Jefferson Laboratory (JLab) \cite{GEp3} and may show a zero crossing for this ratio, if the linear extrapolation of the fit, eq. (\ref{eq:polar}) will be confirmed by the final results.

As no experimental bias has been found in the experiments, the discrepancy between FFs determined from polarized and unpolarized measurements, has been attributed to radiative corrections, as two photon exchange (TPE) \cite{tg1,tg2,tg3,tg4,tg5,tg6} or higher order corrections \cite{By07}. 

\subsection{Time-like region}

The TL region, where $q^2>0$, can be investigated using the crossed reactions $\bar p+p\leftrightarrow e^- + e^+$. Due to unitarity, hadron FFs are complex functions of $q^2$, and their full determination requires more observables as shown in \cite{Zi62,Bi93}, and recently discussed in \cite{ETG05}. However the unpolarized cross section depends only on their moduli, and their measurement is, in principle, simpler than in SL region. In SL region, the Rosenbluth separation requires at least two measurements at fixed $q^2$ and different angles, which implies a change of incident energy and scattered electron angle at each $q^2$ point. In TL region, the individual determination of $|G_E|$ and $|G_M|$ requires the measurement of the angular distribution of the outgoing leptons, at fixed total energy $s=q^2$. Previous experiments (see \cite{An03}), have measured the cross section up to $q^2$=18 (GeV/c)$^2$ and extracted $|G_M|$ in the hypothesis $G_E=G_M$ or $G_E=0$ (which affects up to $30\%$  the values of $|G_M|$). 

Attempts to determine the  ratio ${\cal R}= |G_E|/|G_M|$ can be found in the literature, in ref. \cite{Ba94} (PS170 at LEAR) and more recently in ref. \cite{Babar}, through measurements of the initial state radiation reaction 
(ISR) $e^-+e^+\to \overline{p}+p+\gamma$ (BABAR Collaboration). The results of the two experiments, although affected by large errors, seem to show a different trend. In the second case a larger value was found, in a wide $q^2$ range above threshold. 

The detector $\rm \overline{P}ANDA$, using the antiproton beam plan\-ned at FAIR, will open a new  opportunity to measure TL FFs. The aim of this paper is to show the precision that can be achieved in the measurements of TL proton FFs at \PANDA in a wide range of $q^2$. The interest in spanning a large kinematical domain is the investigation of the transition region from soft to hard scattering mechanisms, which is the domain of perturbative Quantum Chromodynamics (pQCD), where the nucleon can be described in terms of quark and gluon degrees of freedom. In such region, scaling laws and helicity conservation \cite{Ma73,Br73} give predictions for the asymptotic behavior of FFs.

Moreover, the comparison of SL and TL data, allows to verify asymptotic properties which hold for analytical functions \cite{ETG01,ETG05a}. Following the Phragm\`en-Lindel\"of theorem \cite{Ti39}, FFs in TL and SL region have to coincide for $|q^2|\to \infty $. This implies not only that the moduli should be the same, a feature that will be tested with $\rm \overline{P}ANDA$, but also that the phases of TL FFs, which can be accessed only through polarization measurements \cite{Zi62,Bi93,ETG05}, should vanish.
\section{Simulation studies}
\label{sec:Simul}

\subsection{Differential cross section and counting rate}
\label{sec:Diff}

The differential cross section for the annihilation process (\ref{eq:eq1}), first obtained in ref. \cite{Zi62}, is expressed as a function of the proton electromagnetic FFs as:
\ba
\displaystyle\frac{d\sigma}{d(cos\theta)} &=& \displaystyle\frac{\pi\alpha_e^2}{8M^2\tau\sqrt{\tau(\tau-1)}}
\left [ \tau |G_M|^2(1+\cos^2\theta)+ \right .\nn\\
&&\left . |G_E|^2\sin^2\theta
\right ], 
\label{eq:eq2}
\ea
where $\theta$ is the electron production angle in the center of mass system (CM). The $\cos^2\theta$ dependence of eq. (\ref{eq:eq2}) results directly from the assumption of OPE, where the spin of the photon is equal to one and the electromagnetic hadron interaction satisfies $C$ invariance. This corresponds, by crossing symmetry, to the linear Rosenbluth $\cot^2(\theta_e/2)$ dependence \cite{Re99}.

The total cross section is:
\be
\sigma=\frac{\pi \alpha_e^2}{6M^2\tau\sqrt{\tau(\tau -1)}}\left( 2\tau |G_M|^2+|G_E|^2 \right ).
\label{eq:eqstot}
\ee

The evaluations of the cross section and of the counting rate require the knowledge of the FFs. For the numerical estimates below, we use a parameterization of $|G_M|$ from \cite{ETG01}, where the numerator is a constant fitted on TL data:
\be
|G_M| ={22.5}\left (1+q^2/0.71\right )^{-2}{\left(1+q^2/3.6\right)^{-1}}. 
\label{eq:GM}
\ee
Here $q^{2}$ is expressed in (GeV/c)$^{2}$. Eq. (\ref{eq:GM}) gives a conservative estimation of the yield at large $q^2$. As the TL $|G_M|$ values have been extracted from cross section measurements assuming $|G_E|=|G_M|$, the same hypothesis is taken for counting rate estimates, on the basis of eq. 
(\ref{eq:eqstot}).
The evaluations of the cross section and of the counting rate have been also performed using the following QCD inspired parameterization of $|G_{E,M}|$, based on analytical extension of the dipole formula eq. (\ref{eq:dipole}) in TL region, where $Q^{2}$ is replaced by $q^{2}$. Corrections based on dispersion relations have been suggested in  \cite{Sh97} to avoid 'ghost' poles in $\alpha_s$ (the strong interaction running constant), and can be included in the following form:
\be
|G_{E,M}^{QCD}|=\frac {\cal D}{s^2\left [\log^2(s/\Lambda^2)+\pi^2 \right ]},~
{\cal D}=89.45~ [\mbox{GeV/c}]^4.
\label{eq:eqqcdbis}
\ee
where ${\cal D}$ is obtained fitting the experimental data and $\Lambda=0.3$ GeV is the QCD scale parameter. The calculated cross section $\sigma$($\sigma_{QCD}$) and the number of counts $N$($N_{QCD}$) are given in table \ref{table:kin}, assuming an integrated luminosity of ${\cal L}=2$ fb$^{-1}$, which is expected for each data point in four months data taking, with 100 $\%$ efficiency and full acceptance\footnote{This value will be always used below, except when explicitely indicated.}. It is assumed that $|G_E|=|G_M|$, calculated from Eqs. (\ref{eq:GM}) and (\ref{eq:eqqcdbis}), respectively.
\begin{table}[h]
\begin{center}
\begin{tabular}{cccccc}
\hline\noalign{\smallskip}
$s$        & $p$ & $\sigma$ & $N$ &$\sigma_{QCD}$ & $N_{QCD}$\\
\mbox{[GeV/c]}$^2$  & \mbox{[GeV/c]}    & \mbox{[pb]}   &     &     \mbox{[pb]}        &\\
\noalign{\smallskip}\hline\noalign{\smallskip}
5.40  &  1.7  &   538 &  1.1 $10^{6}$  &  481 &  9.6 $10^{5}$  \\  
7.27  &  2.78  &  72  &  1.4 $10^{5}$  &  69  &  1.4 $10^{5}$  \\  
8.21  &  3.3  &   32  &  6.4 $10^{4}$  &   33 &  6.5 $10^{4}$ \\ 
11.0  &  4.9  &   4.52&  9.1 $10^{3}$  &  5.48 &  1.1 $10^{4}$  \\  
12.9  &  5.9  &   1.6  &  3.2 $10^{3}$  &    2  &  4.3 $10^{3}$  \\
13.8  &  6.4  &  1   &  2   $10^{3}$    & 1.4 &  2.8 $10^{3}$  \\  
16.7 &  7.9  &  0.29  &  580           & 0.49 &  979    \\
22.3 &  10.9 &  0.04  &  81            & 0.09 &  183   \\
27.9  &  13.4 &  0.01 &  18            & 0.03 &  51  \\
\noalign{\smallskip}\hline
\end{tabular}
\caption{Cross section $\sigma$ ($\sigma_{QCD}$) and number of counts,  $N$  ($N_{QCD}$) from eq. \protect\ref{eq:GM} (eq. \protect\ref{eq:eqqcdbis}) corresponding to an integrated luminosity of ${\cal L}=2$ fb$^{-1}$, for different values of $q^2=s$ and of the antiproton momentum, $p$.} 
\label{table:kin}
\end{center}
\end{table}
The event generator for the reaction (\ref{eq:eq1}), is based on the angular distributions from eq. (\ref{eq:eq2}), with prescription (\ref{eq:GM}) for the magnetic form factor $G_M$. 

\begin{figure}
\begin{center}
\resizebox{0.50\textwidth}{!}{%
 \includegraphics{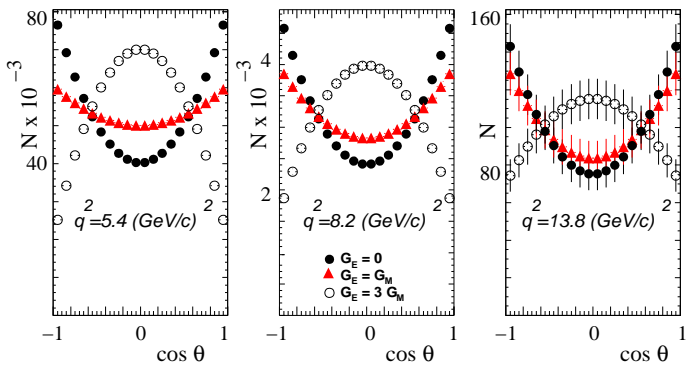}
}
\caption{(Color online) CM angular distributions from the event generator, at $q^2$=5.4, 8.2, and 13.8 (GeV/c)$^2$, for $\bar p+p\to e^-e^+$ and three different hypothesis: ${\cal R}=0$ (black solid circles), ${\cal R}=1$ (red triangles), and ${\cal R}=3$ (black open circles), keeping the same value of the total cross section. }
\label{Fig:distri_ang_GMS}       
\end{center}
\end{figure}

Three different hypothesis were taken for $G_E$. Besides the case $|G_E|=|G_M|$, $({\cal R}=1)$, which is strictly valid only at threshold, the case ${\cal R}=0$ and the case ${\cal R}=3$ (as suggested in ref. \cite{Ba06}), were also considered. The corresponding angular distributions were built keeping the same total cross section at each $q^2$. They are shown in fig. \ref{Fig:distri_ang_GMS}, for three values of $q^2=5.4$, 8.2, 13.8 (GeV/c)$^2$. The reported error bars are statistical only. The sensitivity to ${\cal R}$ decreases when $q^2$ increases, due the falling of the cross section and to the relative weight of the magnetic term, which is growing as $q^2$.

\subsection{Detector description}
\label{detector}
An extensive description of the \PANDA detector and of its different components can be found in ref. \cite{PANDA}. In the present work, we mention some of the characteristics which are important for the specific processes of interest. The conception of the detector, the read out and the acquisition benefit from the experience gained from the construction of recent detectors, such as ATLAS, CMS, COMPASS and BABAR.  The high quality antiproton beam of momentum from 1.5 to 15 GeV/c will be provided by the High Energy Storage Ring (HESR), equipped with electron and stochastic cooling systems. The \PANDA detector should ensure the detection of 2$\cdot 10^7$ interactions/s, with $4\pi$ acceptance and momentum resolution for charged particles at a few percent level. The expected average luminosity ${\cal L}= 1.6 \cdot 10^{32}$ cm$^{-2}$ s$^{-1}$ will be reached with a pellet target of thickness $4\cdot 10^{15}$ hydrogen atoms/cm$^{2}$, and $10^{11}$ stored antiprotons in HESR. The target will be surrounded by a spectrometer which includes a 2 T solenoid magnet. A forward spectrometer, based on a 2 Tm dipole magnet, ensures the detection at small angles, down to $2^{\circ}$. Each spectrometer is equipped with detectors for tracking, charged particle identification, electromagnetic calorimetry and muon identification. The interaction point is reconstructed with the help of a microvertex detector (MVD), consisting of layers of radiation hard silicon pixel detectors surrounded by silicon strip detectors. In addition, charged particle tracking and identification is provided by straw tubes (STT) or a time projection chamber (TPC), completed by GEM detectors at forward angles. The identification of hadrons and leptons in a wide kinematical range requires other complementary detectors. The time-of-flight of particles emitted at large polar angles will be measured in a good resolution time-of-flight barrel. The electromagnetic calorimeter (based on PbWO4 crystals) will provide good energy and time resolution for the detection of photons and electrons at intermediate
energy from a few MeV to $\simeq$ 10 GeV. Detectors based on Cherenkov light (DIRC), which are very efficient for pion-electron separation for momentum $p <1$ GeV/c, will be used in a barrel detector and a forward endcap detector.



\subsection{Simulation, digitization and reconstruction}
\label{sec:Pid}

The simulations are based on the same \PANDA software and the same detector geometry as the one used for the physics benchmark simulations presented in \cite{PANDA}, with STT as central tracker and a pellet target. 
The simulation consists of two steps. The first one, based on the GEANT4 code, is the propagation of the particles through the detector. 
The information on the hits and the energy losses has been digitized, including a model for electronic noise, into a response of the different detectors, in view of using the same Monte Carlo code in simulations and in future real data analysis.

The second step is the reconstruction of the physical quantities particularly important for electron identification such as  momentum, ratio of energy loss to path length in each straw tube $(dE/dx)$, Cerenkov angle in the DIRC detector, and energy deposit in the electromagnetic calorimeter.
 
These two steps have been described in detail in \cite{PANDA} and we will give here only the details which are the most important for the electron identification. 
 
The parameters used to simulate the fluctuations of the physical signals have a direct influence on the particle identification capabilities. Precise estimate  of energy loss fluctuations in thin layers are needed for the central tracker, and were included according to the PAI model for the description of the ionization process \cite{Ap00}.

The $dE/dx$ values are then used for particle identification using the  truncated arithmetic mean method in order to exclude from the sample the largest values corresponding to the extended Landau tail. A truncation parameter corresponding to 70$\%$ out of the $N$ individual $dE/dx$ values was taken to calculate the arithmetic mean, as a compromise between the requirements of the best resolution, defined as the width of the gaussian fit, and the smallest tail of the distribution. A resolution of $<10\%$ is obtained for pions at 1 GeV/c, which corresponds in average to a value of $4\sigma$ of the distance between the two 
truncated means for electrons and pions. 
\begin{figure}
\begin{center}
\resizebox{0.50\textwidth}{!}{%
 \includegraphics{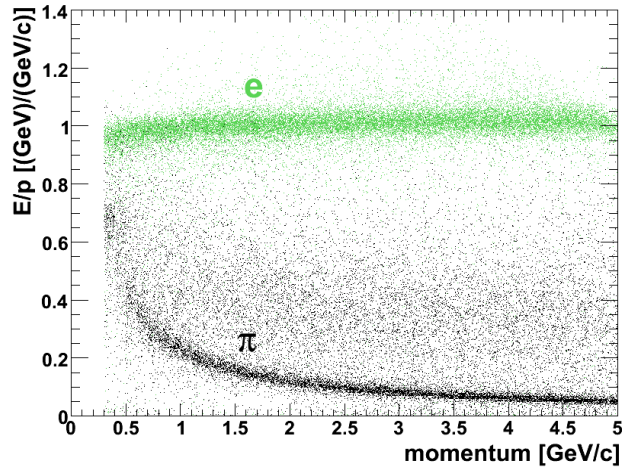}
}
\caption{(Color online) Ratio of the energy deposit in the electromagnetic calorimeter to the particle momentum $E/p$ as a function of $p$, for pions and electrons separation.}
\label{Fig:sep}       
\end{center}
\end{figure}
For the DIRC, the resolution on the Cerenkov angle is given by 
$$ \sigma_{C}=\frac{\sigma_{C,\gamma}}{\sqrt {N_{ph}}}, $$ 
with the single photon resolution $\sigma_{C,\gamma}=10$ mrad. The number of detected photons N$_{ph}$ depends on the velocity and path length of the particle within the radiator and takes into account transmission and reflectivity losses as well as the quantum efficiency of the photodetectors. A nearly Gaussian resolution of about 2.3 mrad is obtained for 1 GeV/c  pions \cite{Fo08}. As the Cerenkov angles for pions and electrons differ by 36 mrad at 500 MeV/c and by 4 mrad at 1.5 GeV/c, the DIRC has a significant discrimination power at the lowest energies.

The electromagnetic calorimeter is the most important detector for the electron identification through the ratio $E/p$ of the measured energy deposit to the reconstructed momentum (see fig. \ref{Fig:sep}). Electrons deposit all the energy in an electromagnetic shower, while muons and hadrons loose only a much lower fraction of their kinetic energy by ionization processes. However, high energy deposits may be due to hadronic interactions within the crystals. In particular, for charged pions undergoing quasi elastic charge exchange processes almost all the energy is transferred to a neutral pion decaying into two photons. These processes are  taken into account by choosing a GEANT4 physics list which includes the Bertini intra-nuclear Cascade model for hadron interactions at intermediate nuclear energies from hundreds of MeV to GeV \cite{He03}.

The shower shape can also be used for particle identification. Due to the small Moli\`ere radius (2 cm) of PbWO4, of the order of the crystal size (2.2 cm), the largest fraction of the electromagnetic shower is contained in a few modules, while a hadronic shower with similar energy is more spread.  The shower shape is characterized by the energy deposits in the central crystal and in the 3x3 and 5x5 module arrays containing the central scintillator. In addition,  a set of four Zernike moments \cite{Zernike} describes the energy distribution within the shower by polynomials which are functions of radial and angular coordinates.

Sets of particles of different species have been generated at given momenta and polar angle. The probabilities for identification of a given particle as electron, muon pion, kaon, or proton are then calculated for each detector using, in addition to the  variables discussed above, the $dE/dx$ information in the microvertex detector, and the information from the muon detector. For the electromagnetic calorimeter, the probability is calculated from the output of a neural network using as inputs  all the parameters of the shower listed above, as described in \cite{PANDA}. As an example, the probability for a pion to be identified as an electron is of the order of $10^{-3}$ at $p$=3 GeV/c, which agrees with values measured on existing detectors (BABAR). From the individual subdetector likelihoods, a global PID likelihood is then calculated. Depending on the signal and background channels, the thresholds can be adjusted in order to ensure the required purity while keeping the signal efficiency as high as possible. In our case, as the pion yield is much larger than the electron one, the threshold is defined by the purity requirement on the electron signal.

\subsection{Study of background channels }

Reactions involving two or more hadrons in the final channel constitute an important background for the measurement of channels with creation of a lepton pair. 

Due to the hermiticity of the detector and the good tracking resolution,  channels with three or more hadrons in the final state will be very efficiently identified. The cross section for channels involving three pions is known to be at most an order of magnitude larger than two pion production. Our simulation show that a reasonable cut on the missing mass gives a rejection factor of at least a factor of hundred.

The larger background is then expected to come from annihilation into two hadrons as $\bar p +p\to\pi^0+\pi^0$, $\bar p +p\to\pi^-+\pi^+$ or $\bar p + p\to K^-+K^+$. The cross sections for the neutral (charged) channels production  are about five (six) orders of magnitude larger than for reaction (\ref{eq:eq1}). In the case of the $\pi^0+\pi^0$ production, $e^-e^+$ pairs are produced  after conversion of the photons from the main $\pi^0$ decay, in particular in the beam pipe before the tracking 
system. In addition, one (or both) $\pi^0$ may undergo Dalitz decay, $\pi^0\to\ e^-+e^++\gamma$, with probability $10^{-2}$ ($10^{-4}$).

In case of charged hadron pair production, both hadrons can be misidentified as leptons. In case of kaon production, the probability of misidentification is lower and kinematical constraints are more efficient, due to their larger mass. 

Therefore, the background coming from $\bar p p$ annihilation into two pions is expected to be the largest and has been evaluated using detailed simulations.
\subsubsection{Simulations of $\bar p + p\to\pi^-+\pi^+$ and $\bar p + p\to\pi^0+\pi^0$ reactions}

\begin{figure}
\begin{center}
\resizebox{0.44\textwidth}{!}{%
\includegraphics{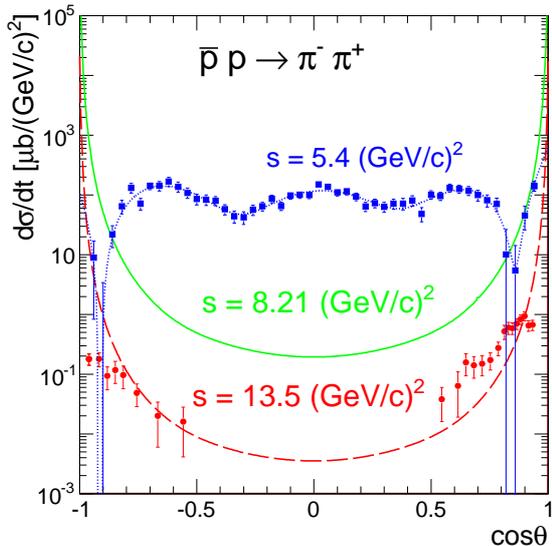}}
\caption{(Color online) CM angular distributions used to simulate the $\bar p+  p\to\pi^-+\pi^+$ reaction for $s$=5.4 (GeV/c)$^2$ (blue dotted line) for $s$=8.21 (GeV/c)$^2$ (green solid line) and $s$=13.5 (GeV/c)$^2$ (red dashed line) as a function of the cosine of the $\pi^-$ production angle. Data are from ref. \protect\cite{pichbe} (squares) and from ref. \protect\cite{Be78,Bu76} (circles).}
\label{Fig:pich}       
\end{center}
\end{figure}

The angular distributions for charged and neutral pion pair production  were extrapolated from a parameterization of the data \cite{pichbe,Be78,Bu76,Ba92,Du78,Ar97}. For $s<6$ (GeV/c)$^2$, the existing data \cite{pichbe,Ba92,Du78} were fitted by Legendre polynomials. In the high energy range, instead, the behavior of exclusive processes is driven by dimensional counting rules, thus the differential cross section of the $\bar p + p\to\pi^-+\pi^+$ process can be parametrized as \cite{Ma73,Br73}:
\be
\frac{d\sigma}{dt}=Cs^{-8}f(\theta)
\label{eq:eqsc}
\ee
where $\theta$ is the CM angle of the $\pi^-$, $t$ is the Mandelstam variable and the function $f(\theta)$ depends on the reaction mechanism. In the framework of the quark interchange dominance model \cite{Gun73}, one has 
\be
f(\theta)=\frac{1}{2}(1-z^2)[2(1-z)^{-2}+(1+z)^{-2}]^{2},~z=\cos\theta.
\label{eq:eqfz}
\ee
$C=440 $ mb (GeV/c)$^{14}$ is a constant, which can not be predicted by QCD, and it is determined from $\pi^+p$ elastic scattering at momentum 10 GeV/c and $\cos\theta$=0. The model predictions were symmetrized ($d\sigma(\theta)/dt\to [d\sigma(\theta)/dt+d\sigma(\pi-\theta)/dt]/2$) and readjusted in the region around 90$^{\circ}$, at each $s$ value, to get a better agreement with the data. The results of the event generator are shown in 
fig. \ref{Fig:pich} for $s$=5.4  (GeV/c)$^2$, $s$=8.21 (GeV/c)$^2$ and $s$=13.5 (GeV/c)$^2$ and compared to data obtained in refs. \cite{pichbe,Be78,Bu76}.

\begin{figure}
\begin{center}
\resizebox{0.45\textwidth}{!}{%
\includegraphics{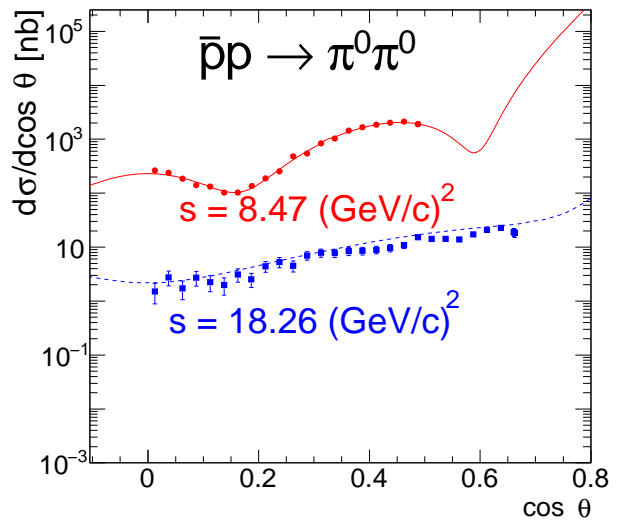}}
\caption{(Color online) Angular distribution of pions produced in $\bar p+p \to\pi^0+\pi^0$ annihilation, for $s=8.47$ (GeV/c)$^2$ (red circles) and $s= 18.26$ (GeV/c)$^2$ (blue squares). Data are from ref. \protect\cite{Ar97}. }
\label{Fig:pi0}       
\end{center}
\end{figure}

For exclusive $\pi^0\pi^0$ production at high energy the following parameterization was taken:
\be
\frac{d\sigma}{d\cos\theta}=\frac{f(s,\theta)}{s^6(\sqrt{tu}/s)^4},~f(s,\theta)=\sum_ia_i(s)P_i(\cos\theta)
\label{eq:eqpi0}
\ee
where $P_i(\cos\theta)$ are Legendre polynomials and are fitted to the data  from E760 at Fermilab, in the kinematical range  $8.5<s<18.3 $ GeV/c \cite{Ar97}, as shown on fig.~\ref{Fig:pi0}. The quality of the fits can be seen in fig. (\ref{Fig:pi0}), where examples of differential cross sections for $\bar p + p\to \pi^0+\pi^0$ are shown. 

For both reactions, the extrapolation to $|\cos\theta|=1$ is affected by a large uncertainty, as no data exist at high energy. Therefore, in the following, only the angular region $ |\cos\theta|\le 0.8$ will be considered, in particular for the evaluation of statistical and systematic errors\footnote{Note that also the electron identification efficiency becomes very low above $ |\cos\theta |= $0.8.}.
In this region, the ratio of $\pi^-+\pi^+$ to $e^-+e^+$ cross sections varies from $10^5$ at $ |\cos\theta|=0 $ to $3\cdot 10^6$ at $|\cos\theta|=0.8 $. On the basis of these numbers, the rejection power should be larger than $3 \cdot 10^9$ ($3\cdot 10^{8}$) for $\pi^-+\pi^+$ ($\pi^0+\pi^0$) in this angular range to limit the background to 0.1$\%$ level.

In order to evaluate the  background rate fulfilling the $e^-+e^+$ criteria, the generated $\pi^-+\pi^+$ and $\pi^0+\pi^0$  events were analyzed using the same PID cuts and kinematical constraints as for the signal.

\subsubsection{Suppression of  $\pi^-\pi^+$ and $\pi^0\pi^0$ backgrounds}

Due to the difference of six order of magnitude in the cross section, between the signal and the $\pi^-\pi^+$ background, an event sample of at least $10^8$ $\pi^-\pi^+$ events was simulated at $q^2$=8.21, 13.8, and 16.7 (GeV/c)$^2$. 

To discriminate pions from electrons, cuts have been applied to the PID combined likelihood for the assumption that the detected particle is an electron. The numbers of simulated $\pi^-\pi^+$ events left after Loose, Tight and Very Tight PID cuts (corresponding respectively to minimum values of the identification probability 85$\%$, 99$\%$ and 99.8$\%$ for each lepton of the pair) are displayed in table \ref{tab_sim}. It is shown that the Very Tight cuts are needed to reach a rejection factor of a few 10$^7$.

Further selection based on the reaction vertex and on the kinematical fit method is applied. The kinematical fit method is a constrained fit, which takes into account energy and momentum conservation. From this, a confidence level (CL) associated to $\pi^+\pi^-$ hypothesis, $CL_{\pi}$ and a second one, $CL_{e}$ corresponding to $e^+e^-$ hypothesis are calculated. The selection of the electrons results from two conditions: $CL_{\pi,e}>10^{-3}$ (which corresponds to $\chi^2<7$ for the kinematical fit) and $CL_{e}> 10~CL_{\pi}$ . These conditions result in an additional rejection factor of the background of $\simeq 100$. Finally, combining the kinematical fit with the PID ends up in a overall background suppression factor of the order of a few $10^9$.

It has been checked that, for $|\cos (\theta)| < 0.8$, the $\pi^-\pi^+$ contamination does not depend drastically on angle and will remain below $ 0.1\%$ in the $q^2$ range of interest.
\begin{table}[h]
\begin{center}
\begin{tabular}{|c|c|c|c|c|}
\hline
\multicolumn{2}{|c|}{$q^2$  [GeV/c]$^2$}      & 8.2 & 12.9 & 16.7 \\
\hline
\multicolumn{2}{|c|}{no cut}               &  $10^8$ & $10^8$     &  2$\cdot 10^{8}$   \\ 
\multirow{3}{*}{ PID cuts}
& Loose      &  425    & 1.2$\cdot10^3$  &  3$\cdot 10^{3}$   \\  
&Tight        &  31    & 70         &  120         \\ 
&Very Tight &  2     & 5          &  6  \\  
\multicolumn{2}{|c|}{kinematic fit(CL)} &  8$\cdot 10^5$ &  $10^6$ &  2.5$\cdot 10^6$  \\
\hline
\end{tabular}
\caption{Number of $\pi^-\pi^+$ events, misidentified as  $e^-e^+$, left after Loose, Tight and Very Tight PID cuts corresponding to respective minimum values  of the
 electron identification probability   85$\%$, 99$\%$ and 99.8$\%$ and  after the confidence level (CL) cut on the kinematic fit for three different 
 $q^2$ values (see text). }
\label{tab_sim}
\end{center}
\end{table}
Concerning the $\pi^0\pi^0$ channel, Dalitz decay, $\pi^0\to\ e^-+e^+ +\gamma$, has a probability $10^{-2}$. Three processes can be sources of $e^-e^+$ pairs: i) double Dalitz decay of the two $\pi^0$, ii) Dalitz decay of one of the pion associated with gamma conversion from the other pion, iii) photon conversion  from two different pions. All these processes, with comparable rates, produce a six particle final state. Thus, even if the produced $e^-e^+\gamma$ pairs fulfill the PID cut, the kinematical constraints give a rejection factor, which combined with the $10^{-4}$ probability for such processes, lead to an efficient suppression of this background. Moreover, by requiring that only a single  $e^-+e^+$ pair has been identified in the whole detector solid angle, it is possible to reduce even further the contribution of this  channel.

\subsection{Analysis of the  $e^-+e^+$ channel}

\begin{figure}
\begin{center}

%
\resizebox{0.45\textwidth}{!}{%
\includegraphics{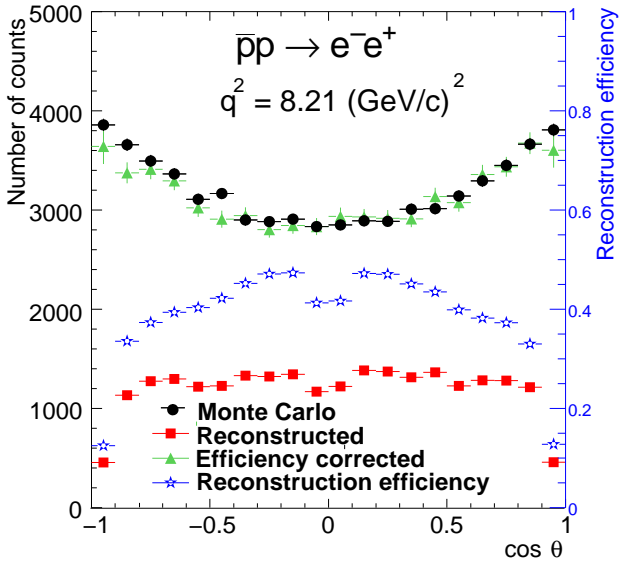}}
\caption{(Color online) Angular distribution of electrons from $e^- e^+$ pairs as a function of $\cos\theta$ at $q^2=8.21$ (GeV/c)$^2$: generated events (black circles), reconstructed events (red squares), acceptance and efficiency correction (blue stars, left scale), and efficiency corrected events (green triangles). }
\label{Fig:fig4.105}       
\end{center}
\end{figure}

Simulations were done for reaction (\ref{eq:eq1}) at $q^2$ values corresponding to table \ref{table:kin}, for ${\cal R}$ =0, 1, and 3. 
A realistic projection of the angular distribution of $e^-e^+$ events, as it will be measured with \PANDA is given in fig.~\ref{Fig:fig4.105}, for $q^2=$ 8.21 (GeV/c)$^2$, and assuming $|G_E|=|G_M|$. The reconstructed events (red squares) are obtained after full Monte Carlo simulation which takes into account tracking, detector efficiency, and acceptance as described in sec.~\ref{sec:Pid}. The reconstruction efficiency corrections have been obtained from an independent simulation which assumes an isotropic electron distribution (blue stars, right scale). Once corrected for this efficiency, the distribution (green triangles) nicely agrees with the 
generated one (black circles). One can see that at this $q^2$ value the average efficiency is of the order of 40$\%$.

The reconstruction efficiency depends on the angle. The sudden drop at $|\cos\theta|>0.8$
corresponds to a decrease of the PID efficiency. At $q^2$=8.21 (GeV/c)$	^2$ the poor $dE/dx$ identification from the STT is responsible of this drop. The loss of efficiency at $\cos\theta=0 $ is due to the target system.

\begin{figure}
\begin{center}
\resizebox{0.45\textwidth}{!}{
\includegraphics{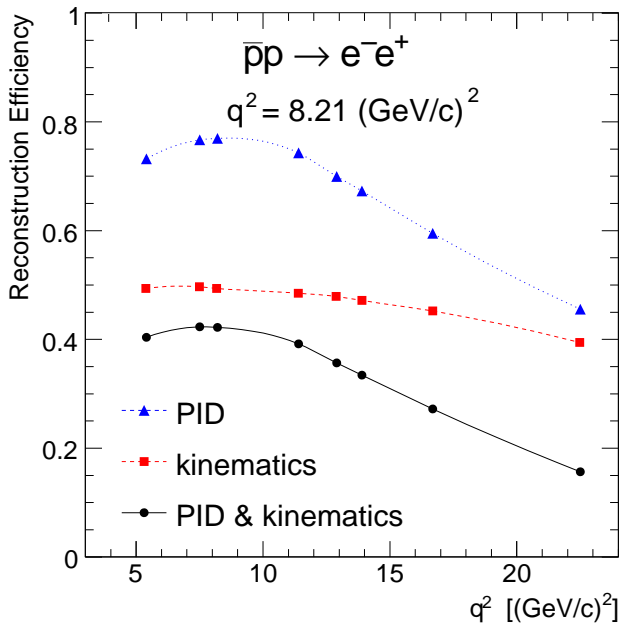}}
\caption{(Color online) Reaction $\bar p +p \to e^+ + e^-$: average reconstruction efficiency as a function of $q^2$ (black solid line). The effect of applying kinematical constraints (red dashed line) and PID cuts (blue dotted line) are separately shown. }
\label{Fig:fig4.104}       
\end{center}
\end{figure}

The  reconstruction efficiency, after integration over the angular range $|\cos\theta|\le 0.8 $, is shown in fig. \ref{Fig:fig4.104}. It is maximum at $q^2\sim 8$ (GeV/c)$^2$ and decreases to 15\% at $q^2 \sim $ 23 (GeV/c)$^2$.
The effects of PID and kinematical constraints are shown separately. The  drop at large $q^2$ is mainly due to PID cuts, as the laboratory angular distribution is more forward peaked with increasing $q^2$, whereas the kinematical selection shows a rather constant behavior. 

The normalization of the measured counting rates will be provided using the $\bar{p}p$ luminosity detector, with an expected precision of 3$\%$. 

Standard radiative corrections are included in the simulation program, via the PHOTOS package \cite{Wa08}. The data will have to be corrected for soft and hard photon emission, which partially compensate each other, reducing the overall effect. Radiative corrections strongly depend on the kinematical conditions and on the criteria for data selection. At $q^2$=9 GeV$^2$  rough estimate gives an overall effect of 10-15\%. Detailed studies of the effects of radiative corrections for this specific channel will be object of a separate paper.

\section{Results and discussion}
\label{sec:Results}

\begin{figure}
\begin{center}
\resizebox{0.50\textwidth}{!}{%
\includegraphics{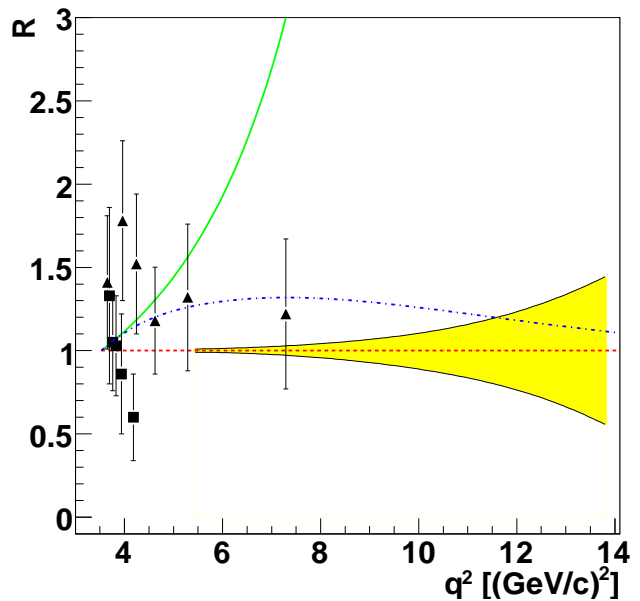}
}
\caption{(Color online) Expected statistical precision on the determination of the ratio ${\cal R}$, (yellow band) for ${\cal R}=1$, as a function of $q^2$, compared with the existing data from Refs. \protect\cite{Babar} (triangles) and \protect\cite{Ba94} (squares). Curves are theoretical predictions (see text).}
\label{Fig:fig4.106}       
\end{center}
\end{figure}

For each $q^2$ value, the simulated differential cross section was fitted with a two-parameter function, in order to extract a global normalization $\alpha$ and the form factor ratio ${\cal R}$, according to:  
\be
N(\cos\theta)=\alpha[\tau(1+\cos^{2}\theta)+ {\cal R}^2\sin^2\theta].
\label{eq:eqfitRs}
\ee
The results are shown in fig. \ref{Fig:fig4.106}, where the expected statistical uncertainty on ${\cal R}$ is plotted as a function of $q^2$ as a yellow band for the case ${\cal R}=1$, and  compared with the existing values from refs. \cite{Ba94} (squares) and \cite{Babar} (triangles). Different methods were used to extract the errors bars for ${\cal R}$. Different kinds of fits where performed with MINUIT, extracting ${\cal R}$ or ${\cal R}^2$ from the quadratic expression (\ref{eq:eqfitRs}) and also from the angular asymmetry which enters linearly in a $\cos^2\theta$ distribution (see sect. 5). A method based on Montecarlo was also used. Detailed discussion and numerical values for the different values of ${\cal R}$ can be found in ref. \cite{Ma08}. As an example, for $q^2$=13.84 (GeV/c)$^2$, one obtains from a Montecarlo approach: 
${\cal R}=1^{+0.54}_{-0.51}$, ${\cal R}=3^{+0.90}_{-0.53}$ and an upper limit of 0.61 (CL=$68\%$) for ${\cal R}=0$. Therefore we concluded that a meaningful value for ${\cal R}$ can be extracted up to $q^2\sim 14$ (GeV/c)$^2$. In the low $q^2$ region, the precision is at least an order of magnitude better than for the existing data. With a precise measurement of the luminosity, this will allow to extract the moduli of $G_E$ and $G_M$, to be compared with the corresponding SL values and with model calculations.

Model predictions display a quite large dispersion, as shown in fig. \ref{Fig:fig4.106}. A QCD inspired parameterization, based on scaling laws \cite{Ma73,Br73}, predicts ${\cal R}=1$, as it depends only on the number of constituent quarks (red dashed line). The green solid line is based on the vector meson dominance (VDM) approach from ref. \cite{Ia73}, and grows up to  $q^2\sim 15$ (GeV/c)$^2$. The blue dash-dotted line is the prediction from ref. \cite{Lo02}, based also on VDM, but including terms to ensure the proper asymptotic behavior predicted by QCD. These models, originally built in the SL region, have been analytically extended to the TL region and the parameters have been readjusted in ref. \cite{ETG05} in order to fit the world data in the whole kinematical region ($i.e.$, in SL region, the electric and magnetic proton and neutron FFs, and in TL region, the magnetic FF of the proton and the few existing data for neutron \cite{Fenice}). Although these models reproduce reasonably well the FFs data, they give very different predictions for the form factor ratio. It is also shown in ref. \cite{ETG05} that polarization observables show large sensitivity to these models. 

\begin{figure}
\begin{center}
\resizebox{0.5\textwidth}{!}{%
  \includegraphics{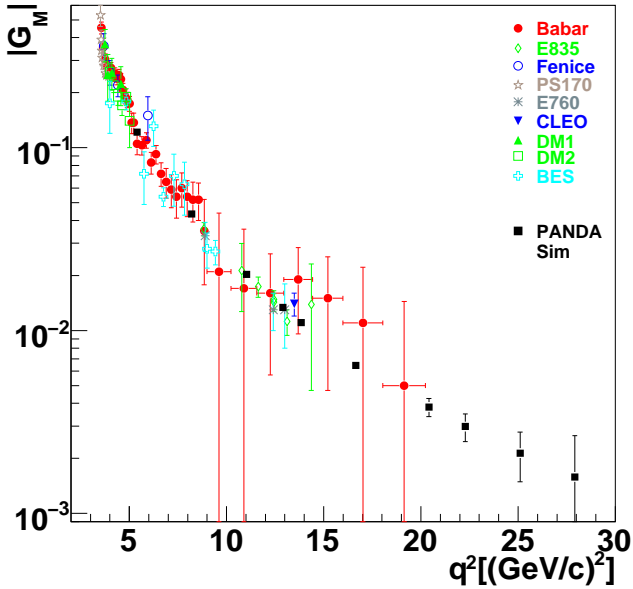}
}
\caption{(Color online) $q^2$ dependence of the world data on the effective proton TL FF, $|G_M|$, as extracted from the annihilation cross section assuming $|G_E|=|G_M|$:  
BABAR \protect\cite{Babar} (red full circles); 
Fenice \protect\cite{Fenice} (blue open circles ); 
E835 \protect\cite{An03,Ambrogiani} (green open lozenge); 
PS170  \protect\cite{Ba94} (gray open stars ); 
E760  \protect\cite{Ar97} (blue asterisk ); 
DM1  \protect\cite{Delcourt} (green full triangles); 
DM2  \protect\cite{Bisello} (green open squares);  
BES  \protect\cite{Ablikim} (cyan open cross ); 
CLEO  \protect\cite{Pedlar}(blue triangle down ); 
\PANDA (full black squares) corresponding to an integrated luminosity of 2 fb$^{-1}$, errors are statistical only (this work).}
\label{Fig:fig4.107}       
\end{center}
\end{figure}

At larger  $q^2$, when the sensitivity of the experiment will make the extraction of ${\cal R}$ meaningless, it will then be possible to extract $|G_M| $ under a definite hypothesis on the ratio, in general ${\cal R}$=1, as done in previous measurements. With a precise knowledge of the luminosity, the absolute cross section can be measured up to $q^2\sim 28$ (GeV/c)$^2$.  The precision of such measurement is shown in fig. \ref{Fig:fig4.107}. 
%
The comparison with the world data shows an expected improvement of at least a factor of ten. Here only the statistical accuracy, based on the number of events measured and identified is taken into account. The reported error bars are  based on conservative extrapolation of the values reported in table \ref{tab_sim}.

Systematic effects of the tracking and reconstruction procedure will be evaluated on the real data, mostly by measurements on known reactions, which are the source of well controlled data samples. 

Comparing such data samples to simulations, as a
function of momentum and angle, will allow to check the rejection
power against pions for each subdetector and to determine the
electron identification efficiency. This insures the control of the 
global rejection (purity of the electron data sample) and the knowledge of the overall
electron reconstruction efficiency. 

The electromagnetic
calorimeter is a crucial ingredient for the electron/pion separation. Besides the  calibration using cosmic rays, several reactions can be identified. The reaction $\bar p +p \to\pi^0 + \pi^0$ can be used for the calibration of all electromagnetic calorimeter crystals at the sub percent level. The two body hadronic reaction $\bar p +p \to \pi^+ + \pi^-$ allows to tune the PID parameters related to pions, and to check the calibration and the resolution of the momentum reconstruction at permille level, due to the high statistics.  Systematic effects from detector misalignments are expected to be negligible. In addition, the decay $J/\Psi \to e^++e^-$, with $6\%$ branching ratio, provides electron data samples of good purity and known angular distribution. It will constitute a concrete measurement of the individual detector responses to electrons. At $q^2$=9.6 (GeV/c)$^2$, one week measurement will be affected by 1\% statistical precision in a 0.1 $\cos\theta$ bin.

\section{Sensitivity to two photon exchange}
\label{sec:twophoton}

As stressed in the introduction, the expression of the cross section (\ref{eq:eq2}) assumes OPE. TPE is suppressed by a factor of $\alpha_e$. At large $q^2$, however, TPE could play a role, due in particular to a  possible enhancement from a mechanism where the momentum is equally shared between the two photons \cite{twof1,twof2,twof3,twof4}. Recently, the possibility of a sizable TPE contribution has been discussed in connection with discrepancies between experimental data, on elastic electron deuteron scattering \cite{Re99} taken in different experiments, and elastic electron proton scattering in polarized and unpolarized experiments \cite{CFP08,tg1,tg2,tg3,tg4,tg5,tg6,Re04a,Re04b,Re04c,Ch07}. Experimentally, no model independent and unambiguous evidence of TPE (more exactly, of the real part of the interference between OPE and TPE) has been found in the experimental data \cite{Ga08,ETG05b,ETG08,Tv05}. Recent analysis of $e^{\pm}p$ cross sections are controversial due to the uncertainties of the data \cite{ETG09,Al09,Ar04}.

The general  analysis of experimental observables in the reaction $\bar p+p\to e^-+e^+$ \cite{Ga06} and in the time reversed channel \cite{Ga05}, taking into account the TPE contribution, was derived in a model independent formalism developed for elastic electron proton scattering \cite{Re04a,Re04b,Re04c}. It was shown that in presence of TPE, the matrix element contains three complex amplitudes: $\widetilde{G}_E$, $\widetilde{G}_M$ and $F_3$, which depend on two kinematical variables, and that the angular distribution contains new terms which are odd in $\cos\theta$ and are of the order of $\alpha_e $ compared to the dominant contribution \cite{Ga06}. 

Therefore, it seems interesting to study the possibility of identifying the TPE effect in the present experiment. The purpose of this study is not to determine the physical amplitudes, but to set a limit for a detectable odd $\cos\theta$ contribution, eventually present in the data. 
As the TPE amplitudes are not known, we used the presence of odd terms in $\cos\theta$ as a (model independent) signature, introducing drastic approximations: we neglected those  contributions to 
$G_{E,M}$ which are smaller by an order $\alpha_e$. We considered only the real part of the three amplitudes, denoted $G_E$, $G_M$ and $F_3$, as their relative phases are not known. 

We approximated the differential cross section in the following way: 
\be\label{eq:eq1G}
\frac{d\sigma}{d\Omega } = \frac{\alpha_e^2}{4q^2}\sqrt{\frac{\tau }
{\tau -1}}D,
\ee
by taking
\ba
\label{eq:eq3a}
D&\simeq&G_M^2(1+\cos^2\theta )+\frac{G_E^2}{\tau }
\sin^2\theta + \nonumber\\ 
&&2\sqrt{\tau (\tau -1)}\left(\frac{G_E}{\tau }-G_M \right )F_3\cos\theta \sin^2\theta.
\ea
Angular distributions were simulated according to eq. (\ref{eq:eq3a}), assuming $G_E=G_M$ for $q^2$=5.4, 8.2 and 13.8 (GeV/c)$^2$ and for $F_3/G_M$=2\%, 5\% and 20\%. A number of events corresponding to Table I were processed for each of these $q^2$ values and the detector efficiency was taken into account. The TPE components induce a distorsion in the angular distributions which vanishes at $\cos\theta$=0 and $\pm$ 1. In order to analyze the distributions, and extract the values of the two photon amplitude, we rewrite the angular distribution as a polynomial in $\cos\theta$. 
In case of OPE, Eq. (\ref{eq:eq2}) can be rewritten as
\be
\displaystyle\frac{d\sigma}{d(\cos\theta)}=
\sigma_0\left [ 1+{\cal A} \cos^2\theta \right ],
\label{eq:eqsa}
\ee
where $\sigma_0$
is the value of the differential cross section at
$\theta=\pi/2$ and ${\cal A}$ is the angular asymmetry \cite{ETG01}.


Therefore, at each value of $q^2$, we can fit the angular distributions with a straight line in $\cos^2\theta$, 
\be
y=a_0+a_1 \cos^2\theta,
\label{eq:eqlin}
\ee 
where $a_0$ and $a_1$ are related to the physical FFs.
Deviations from a straight line are the evidence of the presence of higher order terms, beyond Born approximation. In order to check the sensitivity to odd terms, we fit the angular distributions by the function:
\ba
y=a_0+a_1 \cos^2\theta+ a_2 \cos\theta(1-\cos^2\theta), 
\label{eq:eqa2}
\ea
where $a_2$ is directly related to the ratio $F_3/G_M$.

\begin{figure}
\begin{center}
\resizebox{0.5\textwidth}{!}{%
  \includegraphics{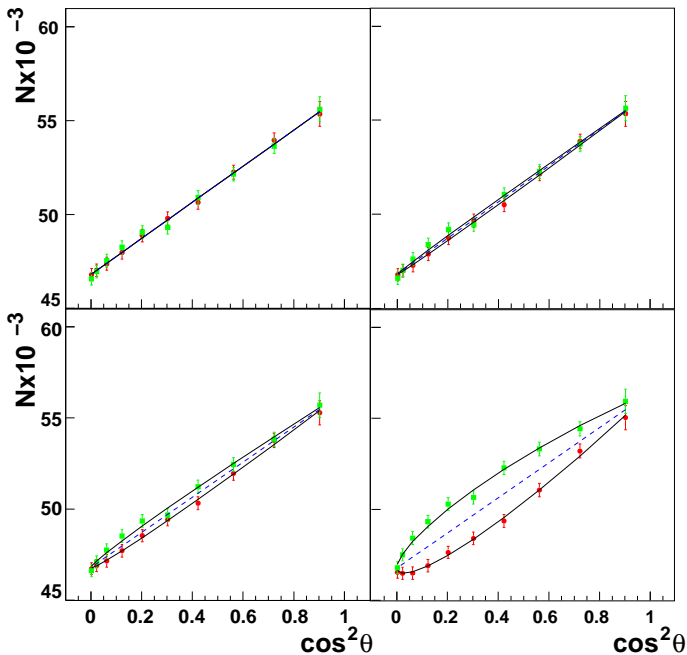}
}
\caption{(Color online) Angular distribution as function of $\cos^2\theta$, according to eq. (\protect\ref{eq:eq3a}), for $q^2$ =5.4 (GeV/c)$^2$ and for different contribution of TPE : no contribution (top left), $2\%$ contribution (top right), $5\%$ contribution (bottom left), and $20\%$ contribution (bottom right). The solid (dashed) line is the result of the fit from eq. (\protect\ref{eq:eqa2}) (from eq. (\protect\ref{eq:eqlin})) . Errors are statistical only.}
\label{Fig:fig3}
\end{center}
\end{figure}

The results of the fit are reported in table \ref{table:res_cubic}. In the case of OPE, as expected, the coefficient $a_2$ is compatible with zero. The odd $\cos\theta$ contribution starts to be visible for $F_3/G_M\ge  5\%$. Note however that the extraction of ${\cal R}$ and ${\cal A}$ is not  affected, in the limit of the error bars, by the presence of the C-odd term and that the $a_0$ and $a_1$ terms are very stable, even at large $q^2$, although the statistical errors are more sizable. Here the error on ${\cal R}$ is derived from ${\cal A}$ by first order derivation. Figure \ref{Fig:fig3} shows the angular distribution as a function of $\cos^2\theta$, for $q^2$=5.4 (GeV/c)$^2$. The lower - red dots (upper - green dots) branches correspond to backward (forward) emission for a negative lepton. The solid (dashed) line is the result of the fit from eq. (\ref{eq:eqa2}) (from eq. (\ref{eq:eqlin})), which includes (does not include) the odd $\cos\theta$ terms.

Due to crossing symmetry properties, the reaction mechanism should be the same in SL and TL regions, at similar values of the transferred momentum. If TPE is the reason of the discrepancy between the polarized and unpolarized FFs measurements in SL region, a contribution of $5\%$ is necessary to bring the data in agreement in the $Q^2$ range between 1 and 6 (GeV/c)$^2$ \cite{CFP08}. The \PANDA simulations show that such level of contribution will be detectable in the annihilation data. We have shown the stability of the extraction of $a_0$ and $a_2$, from the data, even in presence of a relatively large contribution of TPE, in the approximation (\ref{eq:eq3a}). This is due to the symmetry properties of the angular distribution and it has to be taken with caution: the relations  between the observables (differential cross section and angular asymmetry) and FFs hold only in frame of OPE. The signification of the extracted parameters in terms of the moduli of the two electromagnetic FFs is not valid anymore. 

Let us stress that the main advantage of the search of TPE in TL region is that the information is fully contained in the angular distribution (which is equivalent to the charge asymmetry). In the same measurement, the odd terms corresponding to TPE can be singled out, whereas in SL region, in case of TPE, it is necessary to measure electron and positron scattering, in the same kinematical conditions. TPE effects cancel if one does not measure the charge of the outgoing lepton, or in the sum of the cross section at complementary angles, allowing to extract the moduli of the true FFs \cite{Ga06,Ga05}. 

\begin{table*}
\begin{center}
\begin{tabular}{|c|c|c|c|c|c|c|}
\hline
$q^2$  &$F_3/G_M$ & $a_0 $  & $a_1$  &$a_2$   & $|\Delta {\cal R}|$ &${\cal A}\pm \Delta A$  \\ 
\hline
5.4     &  0     &$ 46798 \pm 118$ &$9626 \pm  320$ & $-4\pm 288$ &  
         0.008 & $0.21 \pm 0.01$\\
        &  2\% & $46795 \pm  118$&$9638\pm 321     $ & $358\pm 289$ &
	  0.008 & $0.21 \pm 0.01$ \\
        &  5\% &$ 46794 \pm 118$ & $ 9634 \pm  321$ & $891 \pm  289$ &  
           0.008 &   $0.21 \pm 0.01$ \\    
        &  20\% &$ 46789\pm 118$  & $ 9655 \pm  321$ & $3539 \pm 289$  &
	0.008 &$0.21 \pm 0.01$ \\
\hline
8.2  & $0$ &$ 2832\pm 28$ &  $1127 \pm 82$ &$-45 \pm 72$ &
        0.035 & $0.40 \pm 0.03$\\
     & 2\%&$2859\pm 28$ &  $1057 \pm 82$& $45 \pm 72$ &
        0.035 &$0.37 \pm 0.03$\\
     &  5\% &$ 2857\pm 28$ &  $1065 \pm 82$&  $163 \pm 72$ &
        0.036 & $0.37 \pm 0.03$\\ 
     &  20\% &$ 2863\pm 28$ &  $1042 \pm 82$& $769 \pm 72$ &
      0.035  &$0.36 \pm 0.03$ \\
\hline     
13.84& 0&$ 85\pm 5$ &  $39 \pm 19$& $5 \pm 14$ &
       0.385&  $0.47 \pm 0.23$\\
     & 2\% &$ 86\pm 5$ &  $41 \pm 19$& $12 \pm 14$ &
       0.385&  $0.47 \pm 0.23$\\
     &  5\%& $ 86\pm 5$ &  $41 \pm 19$& $21 \pm 14$ &
       0.385& $0.47 \pm 0.23$\\
     &  20\% &$ 84\pm 5$ & $49 \pm 19$& $65 \pm 14$ & 
       0.382&   $0.57 \pm 0.23$\\
\hline
\end{tabular}
\caption[]{Results from the fit of the angular distributions, for different TPE contributions, according to eq. (\protect\ref{eq:eqa2}), for  $q^2=5.4$, 8.2, and 13.84 (GeV/c)$^2$. The first line at each $q^2$ corresponds to the one photon approximation. }
\label{table:res_cubic}
\end{center}
\end{table*}

\section{Comparison with previous experiments}
\label{sec:Comparison}
As it has been stressed above, the HESR ring will provide a high intensity antiproton beam. This feature, together with the high performance of the \PANDA detector, will allow to have the best measurement until now of FFs in TL region. In table \ref{table:comp} we summarize kinematical and technical aspects of the existing TL FFs experiments. All previous results have been limited by low statistics, which prevented a precise determination of angular distributions. 

In the case of PS170 \cite{Ba94} the detector acceptance was limited, in particular due to the covering in azimuthal angle. The large efficiency for E835 \cite{Ambrogiani} is due to the detection based on gas Cerenkov detector, which gives an average pion suppression factor of the order of $5\cdot 10^{-3}$. This allows to have good identification with relatively loose cuts. In the case of $\rm \overline{P}ANDA$, the quartz-based Cerenkov detector (DIRC) will provide a limited electron pion discrimination, mostly at low momentum (below 1 GeV/c): the necessity to have Very Tight cuts to eliminate the background reduces the electron efficiency. 

The FFs measurement of BABAR \cite{Babar} is indirect, as it is based on initial state radiation. The ISR correction factor due to hard photon emission has to be disentangled, and the angular dependence of the electric and magnetic terms is derived from elaborated simulations. The thorough study of the contributing reactions is described in ref. \cite{Babar}. The final reconstruction efficiency is $\sim17\%$.

\begin{table*}
\begin{center}
\begin{tabular}{|c|c|c|c|c|c|} \hline
Accelerator &  FAIR  & CERN-LEAR & SLAC-PEP II & FERMILAB &  BEPC \\  
		    \hline                    
Experiment &  \PANDA (Sim) & PS170 & BABAR  & E835   &  BES II \\ \hline \hline
Reaction & $\bar p+p\to e^-+e^+\gamma$ &$\bar p+p\to e^-+e^+\gamma$   & $e^-+e^+\to \bar p+p+\gamma $ & $\bar p+p\to e^-+e^+\gamma$   &  $e^-+e^+\to \bar p+p $\\ \hline
$q^{2}$ [GeV/c]$^{2}$ & 5  -  28               & 3.52 - 4.18                  & 3.5 - 20          & 8.84 - 18.4             &  4 - 9.4             \\  
\hline
${\cal L}$ [cm$^{-2}$ s$^{-1}$] & $2 \cdot 10^{32}$     & $3 \cdot 10^{30}$ & $3\cdot 10^{33}$& $2\cdot 10^{31}$     &   $<10^{31}$             \\ 
\hline
$I_{Beam}$  & 10$^{11}$ $\bar{p}/s$   & $3\cdot 10^{6}$ $\bar{p}/s$ & & $5\cdot 10^{11}$ $\bar{p}/s$   &               \\  
\hline
Target   & pellets or jet   & LH2 & collider        & gas jet &  collider \\  
\hline
$|\cos \theta|$     &   $<$0.8    & $<$ 0.8 & $<$ 1 & $<$ 0.62 &  $<$0.8       \\  
\hline
Efficiency &  40$\%$  -  10$\%$  &  $\sim$ 10 $\%$  &  17 $\%$  & $67\%$ &    $\sim 50\%$     \\  
\hline
B/S &  $<1\%$ & $<5\%$   & $<5\%$ &  $<2\%$ & $ 1.5\% - 7.8\%$  \\  \hline
\end{tabular}
\caption{Compared characteristics of TL FFs experiments.}
\label{table:comp}
\end{center}
\end{table*}

\section{Conclusion}
\label{sec:Conclusion}
Feasibility studies of measuring proton TL FFs at \PANDA (FAIR) have been presented. Realistic Monte Carlo simulations, which take into account the geometry, the material budget and the performance of the future detector, as well as tracking efficiency and particle identification have been performed. Background reactions have been studied, with particular attention to two body hadron production. The results show that, applying combined PID criteria and kinematical constraints, it is possible to reach a background/signal ratio of the order of 0.1\%, which is sufficient to ensure a clean identification of $e^-e^+$ pairs corresponding to the reaction of interest (\ref{eq:eq1}).

We have also shown that the reaction $\bar p +p\to e^-+e^+$ at \PANDA will be sensitive to a contribution of TPE of the order of $5\%$ or more with  statistical significance of about two sigma. Note also that systematical errors have not been taken into account. This study profits of one of the main advantages of FFs measurements in the TL region: the angular distribution of the produced electron in one setting contains all the useful information, allowing to extract the true form factors and the TPE contribution as well. 

The cumulated statistics, assuming four months data taking at the nominal luminosity for each $q^2$ value, will give precise information on the proton electric and magnetic FFs, in a wide $q^2$ range. The precision of the ratio of the moduli of the electric and magnetic form factors will be of the order of few percent, in the overlapping region with the data from BABAR, which display errors of the order of 40$\%$. The ratio of the electric to magnetic FF will be measurable until $q^2\simeq 14$ (GeV/c)$^2$, with an error comparable to the existing data taken at much lower $q^2$. Above this value it is still possible to extract a generalized form factor from the total cross section and test its asymptotic and analytic properties.  

The measurement of the cross section allows to access the FFs moduli. In order to determine independently the real and imaginary parts, as well the relative phase, polarization observables are necessary. The possibilities of having a polarized antiproton beam and/or a polarized proton target are under study.

\section{Acknowledgments}
Thanks are due to I. Hrivnacova for help during earlier stages of this work and to S. Pacetti for useful discussions.

%
%

%

\begin{thebibliography}{}
%
%
\bibitem{FAIR} http://www.gsi.de/FAIR. 
\bibitem{PANDA} 
  Physics Performance Report for PANDA: Strong Interaction Studies with
  Antiprotons, The PANDA Collaboration,
  arXiv:0903.3905 [hep-ex]; 
  http://www.gsi.de/PANDA.
\bibitem{ETGnote}
  E.~Tomasi-Gustafsson and M.~P.~Rekalo,
  Int. Report DAPNIA 04-01, arXiv:0810.4245 [hep-ph].
\bibitem{Ro50} M. N. Rosenbluth, Phys. Rev.  \textbf{ 79}, (1950) 615.
\bibitem{Ar86} R. G. Arnold {\it et al.}, Phys. Rev. Lett. \textbf{ 57}, (1986) 174. 
\bibitem{And94} 
L.~Andivahis {\it et al.},
Phys.\ Rev.\ D {\bf 50}, (1994) 5491.
\bibitem{CFP08}
  C.~F.~Perdrisat, V.~Punjabi and M.~Vanderhaeghen,
  Prog.\ Part.\ Nucl.\ Phys.\  \textbf{ 59}, (2007) 694.
\bibitem{Jlab}
  V.~Punjabi {\it et al.},
  Phys.\ Rev.\ C \textbf{71}, (2005) 055202
  [Erratum-ibid.\ C \textbf{71}, (2005) 069902] and refs. therein.
\bibitem{Re68} 
  A.~I.~Akhiezer and M.~P.~Rekalo,
  Sov.\ Phys.\ Dokl.\  \textbf{ 13}, (1968) 572;
  [Dokl.\ Akad.\ Nauk Ser.\ Fiz.\  \textbf{ 180}, (1968) 1081]. 
\bibitem{Re74} 
  A.~I.~Akhiezer and M.~P.~Rekalo,
  Sov.\ J.\ Part.\ Nucl.\  \textbf{ 4}, (1974)  277
  [Fiz.\ Elem.\ Chast.\ Atom.\ Yadra \textbf{ 4}, (1973) 662].
\bibitem{GEp3} C. Perdrisat, Nucl.\ Phys.\  \textbf{ A827}, (2009) 267c.
\bibitem{tg1}
  P.~A.~M.~Guichon and M.~Vanderhaeghen,
  Phys.\ Rev.\ Lett.\  \textbf{91}, (2003) 142303.
\bibitem{tg2}  
  A.~V.~Afanasev, S.~J.~Brodsky, C.~E.~Carlson, Y.~C.~Chen and M.~Vanderhaeghen,
  Phys.\ Rev.\  D \textbf{72}, (2005) 013008.
\bibitem{tg3}
   P.~G.~Blunden, W.~Melnitchouk and J.~A.~Tjon,
   Phys.\ Rev.\  C \textbf{72}, (2005) 034612.
\bibitem{tg4}   
   D.~Borisyuk and A.~Kobushkin,
   Phys.\ Rev.\  C \textbf{74}, (2006) 065203.
\bibitem{tg5}
  C.~E.~Carlson and M.~Vanderhaeghen,
  Ann.\ Rev.\ Nucl.\ Part.\ Sci.\  {\bf 57} (2007) 171;
\bibitem{tg6}
  N.~Kivel and M.~Vanderhaeghen,
  Phys.\ Rev.\ Lett.\  \textbf{103}, (2009) 092004.
\bibitem{By07}
  Yu.~M.~Bystritskiy, E.~A.~Kuraev and E.~Tomasi-Gustafsson,
  Phys.\ Rev.\  C \textbf{ 75}, (2007) 015207.
\bibitem{Zi62}
A. Zichichi, S. M. Berman, N. Cabibbo, R Gatto, 
Nuovo Cim. \textbf{ 24}, (1962) 170.
\bibitem{Bi93}
S. M. Bilenky, C. Giunti, V. Wataghin, Z. Phys. \textbf{ C59}, (1993) 475.
\bibitem{ETG05} 
  E.~Tomasi-Gustafsson, F.~Lacroix, C.~Duterte and G.~I.~Gakh,
Eur.\ Phys.\ J.\ \textbf{  A 24}, (2005) 419.  
\bibitem{An03}
M.~Andreotti {\it et al.},
Phys.\ Lett.\ B \textbf{ 559}, (2003) 20 and refs. therein.
\bibitem{Ba94}
  G.~Bardin {\it et al.},
  Nucl.\ Phys.\  B \textbf{ 411}, (1994)  3.
\bibitem{Babar}
  B.~Aubert {\it et al.}  [BABAR Collaboration],
  Phys.\ Rev.\  D \textbf{ 73}, (2006) 012005.
  \bibitem{Ma73}
V.~A.~Matveev, R.~M.~Muradian, and ~A.~N.~Tavkhelidze, Lett. Nuovo Cim.  \textbf{ 7}, (1973) 719.
 \bibitem{Br73}
S.~J.~Brodsky and G.~R.~Farrar, Phys. Rev. Lett. \textbf{ 31}, (1973) 1153.
\bibitem{ETG01}
  E.~Tomasi-Gustafsson and M.~P.~Rekalo,
  Phys.\ Lett.\  B \textbf{ 504}, (2001) 291.
\bibitem{ETG05a}
  E.~Tomasi-Gustafsson and G.~I.~Gakh,
  Eur.\ Phys.\ J.\  A \textbf{ 26}, (2005) 285.
\bibitem{Ti39} E. C. Titchmarsh, {\it Theory of functions}, Oxford University
Press, London, 1939.
\bibitem{Re99}
  M.~P.~Rekalo, E.~Tomasi-Gustafsson and D.~Prout,
  Phys.\ Rev.\  C \textbf{ 60}, 042202 (1999).
\bibitem{Sh97}
  D.~V.~Shirkov and I.~L.~Solovtsov,
Phys.\ Rev.\  Lett. \textbf{79}, (1997) 1209.
\bibitem{Ba06}
  R.~Baldini, C.~Bini, P.~Gauzzi, M.~Mirazita, M.~Negrini and S.~Pacetti,
  Nucl.\ Phys.\ Proc.\ Suppl.\   \textbf{162}, (2006) 46.
\bibitem{Ap00}
  J.~Apostolakis, S.~Giani, L.~Urban, M.~Maire, A.~V.~Bagulya and V.~M.~Grishin,
  Nucl.\ Instrum.\ Meth.\  A \textbf{453}, (2000) 597.
\bibitem{Fo08}
  K.~F\"ohl {\it et al.},
  Nucl.\ Instrum.\ Meth.\  A \textbf{595}, (2008) 88.
\bibitem{He03}
  A.~Heikkinen, N.~Stepanov and J.~P.~Wellisch,
{\it In the Proceedings of 2003 Conference for Computing in High-Energy and Nuclear Physics (CHEP 03), La Jolla, California, 24-28 Mar 2003}, arXiv:0306008 [nucl-th].
\bibitem{Zernike}
M. Born and E. Wolf, "Principles of Optics", Oxford, Pergamon (1970). 
\bibitem{pichbe} 
E.~Eisenhandler {\it et al.},
  Nucl.\ Phys.\  B\textbf{ 96}, (1975) 109 and refs. therein.
\bibitem{Be78}
  A.~Berglund {\it et al.},
  Nucl.\ Phys.\  B \textbf{ 137}, (1978) 276.
\bibitem{Bu76}
  T.~Buran {\it et al.},
  Nucl.\ Phys.\  B \textbf{ 116}, (1976) 51 and refs. therein.
\bibitem{Ba92} 
  C.~Baglin {\it et al.}  [R704 Collaboration],
  Nucl.\ Phys.\  B \textbf{ 368}, (1992) 175.
 \bibitem{Du78} 
  R.~S.~Dulude {\it et al.},
  Phys.\ Lett.\  B \textbf{ 79}, (1978) 329.
\bibitem{Ar97}
  T.~A.~Armstrong {\it et al.}  [Fermilab E760 Collaboration],
  Phys.\ Rev.\  D \textbf{ 56}, (1997) 2509.
 \bibitem{Gun73} 
  J.~F.~Gunion, S.~J.~Brodsky and R.~Blankenbecler,
  Phys.\ Rev.\  D \textbf{ 8}, (1973) 287.
\bibitem{Wa08}
  Z.~Was, P.~Golonka and G.~Nanava,
  Nucl.\ Phys.\ Proc.\ Suppl.\  \textbf{181}, (2008) 269.
\bibitem{Ma08}  
D. Marchand, T. Hennino, R. Kunne, S. Ong, B. Ramstein, M. Sudol, E. Tomasi-Gustafsson, HAL:in2p3-00374971, IPNO-DR-08-03. 
\bibitem{Ia73}
  F.~Iachello, A.~D.~Jackson and A.~Lande,
  Phys.\ Lett.\  \textbf{ B43}, (1973) 191.
\bibitem{Lo02}
  E.~L.~Lomon,
  Phys.\ Rev.\  \textbf{ C66}, (2002) 045501.
\bibitem{Fenice} 
  A.~Antonelli {\it et al.},
  Nucl.\ Phys.\  B \textbf{ 517}, 3 (1998).
 \bibitem{Ambrogiani}
  M.~Ambrogiani {\it et al.}  [E835 Collaboration],
  Phys.\ Rev.\  D \textbf{ 60}, (1999) 032002.
\bibitem{Delcourt}
  B.~Delcourt {\it et al.},
  Phys.\ Lett.\  B \textbf{ 86}, (1979)  395.
\bibitem{Bisello}
  D.~Bisello {\it et al.},
  Nucl.\ Phys.\  B \textbf{ 224}, (1983) 379.
\bibitem{Ablikim}
  M.~Ablikim {\it et al.}  [BES Collaboration],
  Phys.\ Lett.\  B \textbf{ 630}, (2005)  14.
\bibitem{Pedlar}
  T.~K.~Pedlar {\it et al.}  [CLEO Collaboration],
  Phys.\ Rev.\ Lett.\  \textbf{ 95}, (2005) 261803.
\bibitem{twof1} 
J. Gunion and L. Stodolsky,  Phys. Rev. Lett. \textbf{ 30},
(1973)  345. 
\bibitem{twof2} 
V. Franco,  Phys. Rev. D \textbf{ 8}, 826 (1973).
\bibitem{twof3} 
  V. N. Boitsov, L.A. Kondratyuk and V.B. Kopeliovich, Sov. J.
Nucl. Phys \textbf{ 16}, 287 (1973).
\bibitem{twof4} 
F. M. Lev, Sov. J. Nucl. Phys. \textbf{ 21}, 45 (1973).
\bibitem{Re04a}
M.~P.~Rekalo and E.~Tomasi-Gustafsson,
Eur.  Phys. J. A \textbf{22}, (2004) 331.
\bibitem{Re04b}
M.~P.~Rekalo and E.~Tomasi-Gustafsson,
Nucl.\ Phys.\ A \textbf{740}, (2004) 271. 

\bibitem{Re04c}
M.~P.~Rekalo and E.~Tomasi-Gustafsson,
Nucl.\ Phys.\  A \textbf{742}, 322 (2004).
 
\bibitem{Ch07}
  Y.~C.~Chen, C.~W.~Kao and S.~N.~Yang,
  Phys.\ Lett.\  B \textbf{652}, (2007) 269.
\bibitem{Ga08}
  G.~I.~Gakh and E.~Tomasi--Gustafsson,
  arXiv:0801.4646 [nucl-th], to be published in Nucl. Phys. A.   
\bibitem{ETG05b}
  E.~Tomasi-Gustafsson and G.~I.~Gakh,
  Phys.\ Rev.\  C \textbf{72}, (2005) 015209.
\bibitem{ETG08}
  E.~Tomasi-Gustafsson, E.~A.~Kuraev, S.~Bakmaev and S.~Pacetti,
  Phys. Lett. B \textbf{659}, (2008) 197 .

\bibitem{Tv05}
  V.~Tvaskis, J.~Arrington, M.~E.~Christy, R.~Ent, C.~E.~Keppel, Y.~Liang and G.~Vittorini,
  Phys.\ Rev.\  C {\bf 73}, (2006) 025206.

\bibitem{ETG09}
  E.~Tomasi-Gustafsson, M.~O. Osipenko,  E.~A.~Kuraev, Yu.~Bystritsky and V.~V.~Bytev,
  arXiv:0909.4736 [hep-ph] and refs. therein.

\bibitem{Al09}
  W.~M.~Alberico, S.~M.~Bilenky, C.~Giunti and K.~M.~Graczyk,
  J.\ Phys.\ G  \textbf{36}, (2009) 115009.
\bibitem{Ar04}
  J.~Arrington,
  Phys.\ Rev.\  C \textbf{69} (2004) 032201.

\bibitem{Ga06}
  G.~I.~Gakh and E.~Tomasi-Gustafsson,
  Nucl.\ Phys.\  A \textbf{ 761}, 120 (2005).
\bibitem{Ga05}
  G.~I.~Gakh and E.~Tomasi-Gustafsson,
  Nucl.\ Phys.\ A \textbf{771}, (2006) 169.

  
  \end{thebibliography}
%

\end{document}